\documentclass[3p,times,twocolumn]{elsarticle}
 \biboptions{comma,sort&compress}
 
\usepackage{graphicx}
\usepackage{amsmath}
\usepackage{subcaption}

\usepackage{here}
\usepackage{ecrc}

\volume{00}

\firstpage{1}

\journalname{Nuclear and Particle Physics Proceedings}

\runauth{}

\jid{nppp}

\jnltitlelogo{Nuclear and Particle Physics Proceedings}

\usepackage{amssymb}

\usepackage[figuresright]{rotating}

\usepackage{color}

\begin{document}

\begin{frontmatter}

\title{Importance of fermion loops in $W^+ W^-$ elastic scattering}

\author{Antonio Dobado}
\ead{dobado@fis.ucm.es}   
\author{Carlos Quezada-Calonge\corref{cor0} }  
\ead{cquezada@ucm.es}   
\author{Juan Jos\'e Sanz-Cillero}
\ead{jjsanzcillero@ucm.es}

\cortext[cor0]{Speaker, corresponding author.} 
\cortext[cor1]{Talk given at 23th International Conference in Quantum Chromodynamics (QCD 20, 35th anniversary),  27 - 30 October 2020, Montpellier - FR.}
\address{Departamento de F\'{\i}sica Te\'{o}rica and Instituto IPARCOS,  \\ 
Universidad Complutense de Madrid. Plaza de las Ciencias 1, 28040-Madrid, Spain.} 

\pagestyle{myheadings}
\begin{abstract}
We test the assumption that fermion-loop corrections to high energy $W^+ W^-$ scattering are negligible  when compared to the boson-loop ones. Indeed, we find that, if the couplings of the interactions deviate from their Standard Model values, fermion-loop corrections can in fact become as important or even greater than boson-loop corrections for some particular regions of the parameter space, and both types of loops should be taken into account. 
Some preliminary results are shown.
\end{abstract}
\begin{keyword}  
Effective Theories \sep Chiral Lagrangians \sep Beyond Standard Model \sep Fermion Loops



\end{keyword}

\end{frontmatter}

\section{Introduction}

When testing for new physics (NP), loop corrections must be taken into account to confront   experiments with enough precision. A possible place where NP could be found at the LHC is vector-boson scattering. 
Since no further NP states have been found below the TeV scale, the low-energy interactions between Standard Model particles accept an effective field theory (EFT) description. In particular, if one assumes the existence of a high-energy regime where the electroweak (EW) boson scattering becomes strongly interacting, the Electroweak Chiral Lagrangian (ECL) is the most appropriate EFT approach. In addition, the Equivalence Theorem (ET)~\cite{ET} relates, up to corrections $\mathcal{O}(M_W/ \sqrt{s})$, amplitudes with longitudinal EW gauge bosons and amplitudes with would-be Goldstone-bosons in the external legs. Hence, the ET can be very useful, largely simplifying computations. In this context, because of their formally dominant $\mathcal{O}(s^2)$ scaling with the center of mass energy $\sqrt{s}$, only boson-loop corrections are normally considered; as fermion-loop corrections formally scale like $\mathcal{O}(M_{\rm fer}^2 s)$, they are mostly neglected in previous bibliography. Although this assumption is fair in most cases, we would like to point out in these proceedings this is no longer true for some ranges of values of the effective couplings, where fermion loops turn numerically relevant. 

The expressions for the fermion-loop contributions are proportional to the mass of the fermion  inside the loop and couplings of the Lagrangian. Some of these couplings still allow a 10 $\%$ deviation with respect to their SM values~\cite{higgsxsection}. For this reason it is important to test how relevant are these fermion  contributions when considering the whole range of possible coupling values. In this note we will focus on top quark corrections because they are the most significant for our purposes.
Further details will be given in a forthcoming article~\cite{in-preparation}.

 Our results are obtained by using a Higgs Effective Field Theory (HEFT) or Electroweak Chiral Lagrangian (ECL) equipped with a Higgs field~\cite{Appelquist}. In particular, we focus on the imaginary part because they enter next-to-leading order in the chiral counting and are not masked by the purely real lowest order amplitude. The magnitude of interest will be the ratio fermions/bosons  for the $J=0$ and $J=1$ projected partial-wave amplitudes (PWA).

It is important to note that only longitudinal polarizations of the electroweak bosons have been taken into account.  


\section{Coupling the top quark in the ECL} 

Neglecting the bottom quark mass $M_b \ll M_t$ (in addition to all other SM fermions), the lowest order coupling of the top quark to the HEFT is given by the Yukawa like Lagrangian:
\begin{equation}
\begin{array}{lcl}
         \mathcal{L}_Y & =-\mathcal{G}(h) \bigg [\sqrt{ 1-\frac{\hat{\omega}^2}{v^2}}M_t t \bar{t}+i\frac{\omega ^0}{v} M_t \bar{t}\gamma^5t+ \\ &   -i\frac{ \sqrt{2} \omega ^+}{v}M_t \bar{t} P_L b +
   i\frac{ \sqrt{2} \omega ^-}{v}M_t \bar{b} P_R t \bigg ].
\end{array}
\end{equation}
with $P_{R,L}=\frac{1}{2}(1\pm \gamma_5)$ and where we have introduced the Higgs function:
\begin{equation}
    \mathcal{G}(h)=1+{c_1} \frac{h}{v}+{c_2} \frac{h^2}{v^2}+{c_3}\frac{h^3}{v^3}+{c_4}\frac{h^4}{v^4}+...
\end{equation}

Thus, at leading order (LO), our effective Lagrangian is given by:
\begin{equation}
    \begin{array}{lcl}
\mathcal{L}_S & = \frac{1}{2} \mathcal{F}(h) \partial_\mu \omega ^i \partial^\mu \omega_j \left( \delta_{ij}+\frac{\omega_i \omega_ j}{v^2}\right) + \\ & +\frac{1}{2}  \partial_\mu h \partial^\mu h  - V(h)+ \mathcal{L}_Y.
\end{array}
\end{equation}
with the usual HEFT Higgs function $\mathcal{F}$ multiplying the would-be Goldstone $\omega^a$ kinetic term,
\begin{equation}
    \mathcal{F}(h)=1+2{a} \frac{h}{v}+{b} \frac{h^2}{v^2}+{c} \frac{h^3}{v^3}+...
\end{equation}
and the Higgs potential,
 \begin{equation}
    \mathcal{V}(h)= \frac{M_h^2}{2}  h^2+ d_3 \frac{M_h^2}{2v^2}  h^3+ {d_4} \frac{M_h^2}{8v^2} h^4+...
\end{equation}

In the SM case one has $a=b=1$ and $c=0$, for $\mathcal{F}(h)$, $c_1=1$ and $c_{n \geq 2}=0$ in $\mathcal{G}(h)$, and $d_3=d_4=1$ and $d_{i\geq 5}$ in $\mathcal{V}(h)$. 

For the one-loop elastic $W^+W^-$ scattering discussion in this article the only relevant couplings in the LO Lagrangian will be $a$, $b$, $c_1$ and $d_3$.
Finally notice that for the study beyond the ET, one must also incorporate the EW gauge boson interactions to the ECL~\cite{Appelquist}.

\section{Loop corrections to the elastic $W^+W^-$ scattering}

At LO, $\mathcal{O}(p^2)$, this amplitude $T_2$ is purely real and it is given by tree-level diagrams made from leading order HEFT Lagrangian $\mathcal{L}_2$  vertices. Next contribution $T_4$ shows up at $\mathcal{ O}(p^4)$ and consists of a real tree-level part  $T_{4,\rm tree}$, from the effective couplings in $\mathcal{L}_4$ (namely $a_4$ and $a_5$) and one-loop diagrams coming from  $\mathcal{ L}_2$ vertices giving the $T_{4,1\ell}$ contribution to the $\mathcal{ O}(p^4)$ amplitude showing and imaginary part.

Up to the order studied in this work, $\mathcal{O}(p^4)$, the real part of the amplitude is provided by the mentioned three contributions, $\mbox{Re}T=T_2+T_{4\rm tree}+ \mbox{Re}T_{4,1\ell}$.  
This makes the study of the NLO one-loop corrections cumbersome. On the other hand, the imaginary part only gets contributions from one-loop diagrams up to this order, $\mbox{Im}T=\mbox{Im}T_{4,1\ell}$, and it is fully determined by the LO effective Lagrangian. Therefore, focusing on this imaginary part makes the study of importance of fermion corrections much simpler and clearer, and thus it will be the procedure followed in this note.   
More specifically, we will be studying the imaginary part of the projected partial PWA $a_J(s)$, 
where:
$$
T(s,t)=\sum_J 16\pi K (2J+1) P_J(\cos\theta) \, a_J(s),
$$
with $K=1$ ($K=2$) for distinguishable (indistinguishable) final particles. In the physical energy region, Im $a_J(s)$ will be obtained from the one-loop absorptive cuts in the $s$-channel, which we will use as a measure of the relative importance of the various contributions.

For scattering amplitudes with only bosons as external legs it is possible to clearly separate fermion and boson loop diagrams (no mixed loops appear).  Now we will ponder the relevance of each of these two contributions. For this, we use the following notation to refer to the corresponding absorptive cuts:
\begin{eqnarray}
 {\rm Fer}_J&=&Im[a_J]_{t\bar{t}} \, ,\nonumber\\ 
 {\rm Bos}_J&=&Im[a_J]_{WW,ZZ,hh, \gamma \gamma} \, .
\end{eqnarray}

Previous bibliography provides the various intermediate absorptive cuts: 
longitudinal vector bosons $W^+W^-$~\cite{espriu}, longitudinal vector bosons $ZZ$~\cite{wwzz}, $hh$~\cite{wwhh} and $\gamma \gamma$~\cite{gammagammaww}. For the case of the weak bosons $W^\pm$ and $Z$ only longitudinal polarizations are being considered in these proceedings. This is because we are interested in scenarios with a strongly interacting electroweak symmetry breaking sector where longitudinal components dominate the high energy dynamics. In principle, there is also a contribution from the $Zh$ abortive cut which vanishes in the ET limit and is neglected in this proceedings. 

At this point it is important to notice that the relevant couplings entering in each PWA are:
\begin{equation} \label{eq1}
\begin{array}{llr}
\mbox{\bf J=0:} \qquad 
&{\rm Fer}_0 \quad \longrightarrow\quad  a,c_1 \\
&{\rm Bos}_0 \quad \longrightarrow\quad a,b, d_3 \\
\nonumber 
\\ 
\mbox{\bf J=1:} \qquad 
&{\rm Fer}_1  \quad\longrightarrow\quad  \mbox{no dependence} \\ 
& {\rm Bos}_1 \quad\longrightarrow\quad  a\, .
\end{array}
\end{equation}

The goal of this note is to point out that there are regions of the parameter space where fermion loops become as important as the bosonic loops and thus they should not be neglected. 
In order to explore this possibility we will consider the ratio:
\begin{equation}
    R_J=\frac{{\rm Fer}_J}{{\rm Bos}_J+{\rm Fer}_J} \,\, \in\,\, [0,1]\, ,
\end{equation}
since by unitary the imaginary parts of the PWA are always positive.
Values of $R_J$ close to 0 will indicate that we can safely drop fermion loops while significant deviations from 0 will point out the relevance of the fermion (top quark) contribution  to the  $W^+W^-$ and $hh$ scattering. As it is commonly assumed, we anticipate that fermion loops are negligible in most of the parameter (coupling) space. Nonetheless, we will see that this is not true for some particular channels and in the region around some specific values of the parameters.

\section{Results for elastic $W^+W^-$ scattering}

In this analysis we have explored the couplings in the phenomenological admissible range  \cite{higgsxsection}:

\begin{equation}
0.9\leq \, a,b,c_1 , d_3\,\leq 1.1. \nonumber\\
\end{equation}

Concerning the center-off-mass energy we have considered the interval:  $0.5$~TeV$\leq \sqrt{s}\leq 3$~TeV, which is the relevant one to look for NP at the LHC.

\subsection{$R_0$}

In the  following plots we have scanned this region of the coupling space for different values of the different parameters each time while keeping the others fixed to their SM values for reference. 

At this point it is important to state that, when dealing with values of the parameters close to the SM, the ET  is no longer that useful at the energies considered here. This is because the SM is a renormalizable theory, where the longitudinal components of the electroweak gauge bosons are not strongly interacting and do not play a dominant role~\cite{Delgado:2013loa}. Since the SM would-be Goldstone-boson scattering vanishes in the ET limit $\sqrt{s}\gg M_W$, one has to go beyond that approximation in this case.

\begin{figure}[h]
\begin{subfigure}{1 \columnwidth}
  \centering
  \includegraphics[width=1\linewidth]{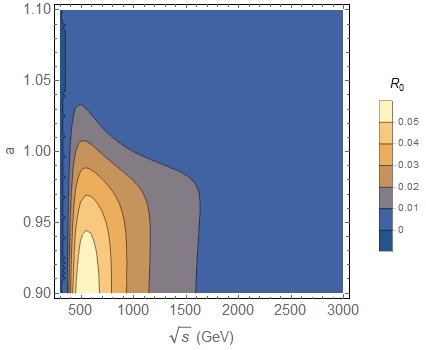}  
  \caption{\small $R_0$ dependence on $a$ for $b=c_1=d_3=1$.\\ \phantom{X}}
  \label{fig:r0chnga}
\end{subfigure}
\begin{subfigure}{1 \columnwidth}
  \centering
  \includegraphics[width=1 \linewidth]{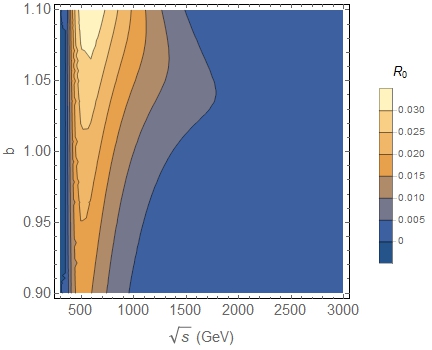}  
  \caption{\small $R_0$ dependence on $b$ for $a=c_1=d_3=1$.}
  \label{fig:r0chngb}
\end{subfigure}
\label{fig:fig1}
\caption{\small}
\end{figure}

As we see in Figure \ref{fig:r0chnga} and Figure \ref{fig:r0chngb}, when we scan $a$ and $b$ we can find corrections of 5 $\%$ around 0.5 TeV. As the energy increases fermion corrections rapidly become irrelevant, as expected.

\begin{figure}[h]
\begin{subfigure}{1 \columnwidth}
  \centering
  \includegraphics[width=\linewidth]{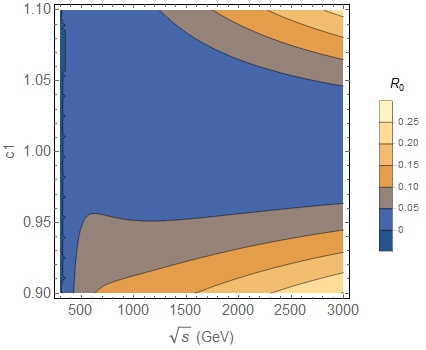}  
  \caption{\small $R_0$ dependence on $c_1$ for $a=b=c_1=1$.\\ \phantom{X}}
  \label{fig:r0chngc1}
\end{subfigure}
\begin{subfigure}{1 \columnwidth}
  \centering
  \includegraphics[width=\linewidth]{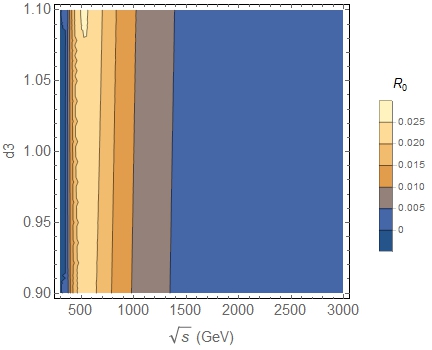}  
  \caption{\small $R_0$ dependence on $d_3$ for $a=b=c_1=1$.}
  \label{fig:r0chngd3}
\end{subfigure}
\label{fig:fig1}
\caption{\small}
\end{figure}

In the  $c_1$ case (Figure \ref{fig:r0chngc1}), we find corrections of about 5\% at 0.5 TeV and as the energy increases we get 25 \%  corrections around 2.5 TeV. The dependence on $d_3$ is negligible as we see in Figure $\ref{fig:r0chngd3}$, with at most 3 \% at low energies.

If we do a parameter scan for all values between 0.9 and 1.1 for two benchmark energies, 1.5 TeV and 3 TeV,  we find that the maximum correction (with $R_0\sim 25\%$) happens for 3 TeV, $a=b=d_3=1$ and $c_1=0.9$. This means that $c_1$ is the most sensitive parameter to fermion corrections . The further $c_1$ is from 1, the bigger the correction is. In this case the $ZZ$ and $hh$ cuts are the most important ones.

\subsection{$R_1$}

\begin{figure}[h]
\includegraphics[width=1 \linewidth]{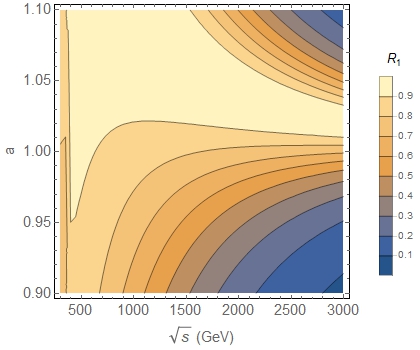}
\centering
\caption{\small $R_1$ dependence on $a$.}
\label{fig:r1chnga}
\end{figure}

For the $J=1$ partial wave, the fermion contribution Fer$_1$ does not depend on any parameter apart of the mass of the top quark, while the boson part Bos$_1$ depends only on $a$ through the $W^+W^-$ cut. As it can be seen in Figure~\ref{fig:r1chnga} we find a wide range of corrections for low energy ( 80-90 \% at 0.5 TeV for $a$ between 0.95 and 1) and for high energies (10-90 \% at 3 TeV in the whole range).

Thus, in this case the assumption that fermion corrections can be neglected does not hold for our study of the imaginary part of the partial waves.

\subsection{Specific Scenarios: Minimal Composite Higgs Model (MCHM)}

From the previous analysis we can see that, when the values of some parameters are below 1, fermion corrections can be in fact relevant. This is the case for example of some NP scenarios like the Minimal Composite Higgs Models (MCHM) where the parameters depend on the NP scale $f$. Choosing a value for $a=a^*$ we can find the scale of new physics $f$ via $a^*=\sqrt{1-\xi}$ with $\xi=v^2/f^2$ \cite{MCHM}. The parameter $b$ follows the expression $b^*=1-2 \xi$ while $c_1^*=d_3^*=a^*$.

\begin{figure}[h]
\begin{subfigure}{1 \columnwidth}
  \centering
  \includegraphics[width=\linewidth]{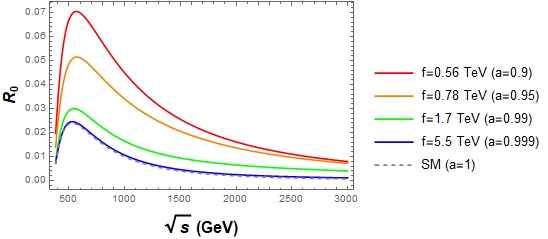}  
  \caption{\small Ratio for the $R_0$ PWA in the MCHM}
  \label{fig:r0composite}
\end{subfigure}
\begin{subfigure}{1 \columnwidth}
  \centering
  \includegraphics[width=\linewidth]{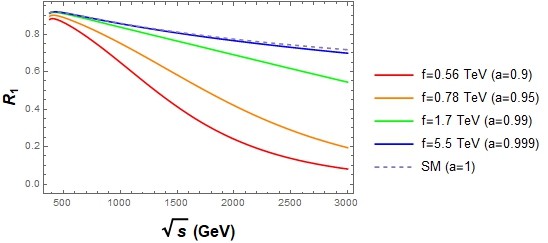}  
  \caption{\small Ratio for the $R_1$ PWA in the MCHM}
  \label{fig:r1composite}
\end{subfigure}
\label{fig:figcomposite}
\caption{\small}
\end{figure}

As seen in Figure \ref{fig:r1composite}, $R_0$ has a maximum of 7 \% for $a=0.9$   around 0.6~TeV and rapidly decreases. For other values closer to the SM corrections are even smaller. For $R_1$ the opposite happens: close to the SM values, corrections are important both at low and high energy. For instance, for $a=0.99$ fermion corrections vary between 90 \% and 60 \%. Again, the second PWA, $a_1$, is more sensitive to fermion loops.

\section{Conclusions}

In the context of HEFT we have studied the fermion loop contribution to the $W^+ W^- \rightarrow W^+ W^-$ amplitude induced by the top quark and compared it with the boson ones. As expected boson contributions dominate in most of the parameter space. However  there are small regions where fermions become relevant. For the $J=1$ partial-wave ratio $R_1$, fermion corrections could be as important, or even greater, than the boson ones. As we showed, the most important parameter for the $J=0$ ratio $R_0$ is $c_1$. In this preliminary analysis we find the largest correction for $a=b=d_3=1$ and $c_1=0.9$ for energies beteween 1.5 and 3 TeV, obtaining a 25\% contribution for the latter. 
%
%
For $R_1$, where we have just the $a$ parameter to play with, we find that, even for values close to the SM, fermion corrections are in fact relevant, going all the way up to 90 \% for $a \sim 1$.

When it comes to the MCHM, the $R_1$ ratio moves from 90 \% (close to the SM) to 20 \% ($a=0.9)$. $R_0$ has a maximum of 7 \% correction at low energies (0.5 GeV) when $a=0.9$. Therefore $R_1$ is the most promising channel to test fermion loop corrections.

In a future work we intend also to  include all the intermediate state polarizations (not only longitudinal) and test their relevance, which were assumed negligible here~\cite{in-preparation}.  Finally, since we are dealing with the possibility of a strongly interacting electroweak symmetry breaking sector, one should have to deal with the problem of unitarity of the whole amplitude~\cite{Dobado:2019fxe}.

\section*{Acknowledgements} We would like to thank our collaborators A. Castillo, R. L. Delgado and F. Llanes-Estrada,  who participated in the earlier parts of the research presented in this note \cite{Castillo:2016erh}. 
This research is partly supported by the Ministerio de Ciencia e Inovaci\'on under research grants FPA2016-75654-C2-1-P and PID2019-108655GB-I00; by the EU STRONG-2020 project under the program H2020-INFRAIA-2018-1 [grantagreement no. 824093]; and by the STSM Grant from COST Action CA16108. C. Quezada-Calonge has been funded by the MINECO (Spain) predoctoral grant
BES-2017-082408.

\end{document}